\documentclass[10pt,conference]{IEEEtran}

\usepackage{etoolbox}
\makeatletter
\def\ps@headings{%
	\def\@oddhead{\mbox{}\scriptsize\rightmark \hfil \thepage}%
	\def\@evenhead{\scriptsize\thepage \hfil \leftmark\mbox{}}%
	\def\@oddfoot{}%
	\def\@evenfoot{}}
\makeatother
\usepackage{verbatim}
\usepackage{xcolor}
\usepackage{amsfonts}
\usepackage{mathrsfs}
\usepackage{amsfonts}
\usepackage{amssymb}
\usepackage{graphicx}
\usepackage{epsfig}
\usepackage{epstopdf}
\usepackage{psfrag}
\usepackage{amsmath}
\usepackage{array}
\usepackage{cases}
\usepackage{subfigure}
\usepackage{cite,graphicx,amsmath,amssymb,color}
\usepackage{algorithmic}
\usepackage{algorithm}
\usepackage{xfrac}
\usepackage{algorithm}
\usepackage{algorithmic}
\usepackage{stmaryrd}
\usepackage{multirow}
\usepackage{subfig}
\usepackage{graphicx,times,amsmath} % Add all your packages here
\DeclareMathOperator*{\argmax}{argmax}

\usepackage{url}
\usepackage{enumerate}
\IEEEoverridecommandlockouts

\newtheorem{lemma}{Lemma}

\newtheorem{problem}{Problem}

\begin{document}
\bibliographystyle{IEEEtran}
	
\title{Optimal Multi-View Video Transmission in OFDMA Systems}
	%\author{Junfeng Guo, \quad  Zhaozhe Song, \quad Ying Cui\\
	%Department of Electronic Engineering, Shanghai Jiao Tong University, Shanghai, P. R. China}

\author{\IEEEauthorblockN{Wei Xu, Ying Cui, Zhi Liu and Haoran Li}\thanks{Manuscript received 
		September~22, 2019; revised November~12, 2019; accepted December~15, 2019.
		The work of Y. Cui was supported in part by NSFC China (61771309, 61671301, 61420106008, 61521062). The work of Z. Liu was supported by JSPS KAKENHI grants 18K18036, 19H04092, and The Telecommunications Advancement Foundation Research Fund. The associate editor coordinating the review of this paper and approving it for publication was Prof. Nkouatchah, Telex M. N. (Corresponding author: Ying Cui.) 
		
		W. Xu and Y. Cui  are with the Department of Electronic Engineering, Shanghai Jiao Tong University, Shanghai 200240, China (e-mail:cuiying@sjtu.edu.cn).
		
		Z. Liu is with the Department of Mathematical and Systems Engineering, Shizuoka  University, Hamamatsu 432-8561, Japan. 
	    
        H. Li is with the School of Computer and Technology, Xidian University, Xi'an 710126, China.}}

\pagestyle{headings}

%Email: \{aaa, bbb, ccc\}@sjtu.edu.cn}
\maketitle
\allowdisplaybreaks[4]

\begin{abstract}
%\boldmath
In this letter, we study the transmission of a multi-view video (MVV) to multiple users in an Orthogonal Frequency Division Multiple Access (OFDMA) system. To maximally improve transmission efficiency, we exploit both natural multicast opportunities and view synthesis-enabled multicast opportunities. First, we establish a communication model for transmission of a MVV to multiple users in an OFDMA system. Then, we formulate the minimization problem of the average weighted sum energy consumption for view transmission and synthesis with respect to view selection and transmission power and subcarrier allocation. The optimization problem is a challenging mixed discrete-continuous optimization problem with huge numbers of variables and constraints. A low-complexity algorithm is proposed to obtain a suboptimal solution. Finally, numerical results further demonstrate the value of view synthesis-enabled multicast opportunities for MVV transmission in OFDMA systems.
\end{abstract}



% \fi
\IEEEpeerreviewmaketitle
\section{Introduction}

A multi-view video (MVV) is produced by simultaneously capturing a scene of interest from different angles with multiple cameras. Depth-Image-Based Rendering (DIBR) can be adopted to synthesize additional views which provide new view angles. Any view angle can be selected freely by an MVV user. MVV has been widely used in education, entertainment, medicine, etc. Compared to a traditional single-view video, an MVV usually has a large size, and hence yields a heavy burden for wireless networks. To improve transmission efficiency, usually, views are encoded separately and only the requested views are transmitted.

In \cite{chen2017Muvi,zhang2019pom}, MVV transmission in single-user wireless networks is studied. The antenna selection and power allocation are optimized to maximize the utility \cite{chen2017Muvi} or minimize the total transmission power\cite{zhang2019pom}. As view synthesis and multicast opportunities are not considered, the proposed solutions in \cite{chen2017Muvi,zhang2019pom} cannot be directly extended to multiuser wireless networks. In \cite{wu2015augmented,zhao2015qos,zhao2014qos,xu2019optimal}, MVV transmission in multiuser wireless networks is studied. To reduce energy consumption, \cite{wu2015augmented,zhao2015qos,zhao2014qos,xu2019optimal} propose transmission mechanisms which can exploit natural multicast opportunities.
In particular, \cite{wu2015augmented,zhao2015qos,zhao2014qos} focus on Orthogonal Frequency Division Multiple Access (OFDMA) systems. The power and subcarrier allocation are optimized to minimize the bandwidth consumption \cite{wu2015augmented} or total transmission power \cite{zhao2015qos,zhao2014qos}. None of \cite{wu2015augmented}, \cite{zhao2015qos} and \cite{zhao2014qos} achieves optimal power and subcarrier allocation, or utilizes view synthesis to further improve transmission efficiency. In \cite{xu2019optimal}, the authors consider Time Division Multiple Access (TDMA) systems and allow view synthesis at the server and users to create multicast opportunities. The total energy consumption minimization with respect to view selection and resource allocation is investigated.

In this letter, our goal is to investigate optimal MVV transmission in OFDMA systems. It yields a more challenging optimization problem than the one in \cite{xu2019optimal} for TDMA systems, as the numbers of variables and constraints grow exponentially with the number of subcarriers and the joint discrete optimization of view selection and subcarrier allocation is highly nontrivial. First, we establish a communication model for transmission of an MVV to multiple users in an OFDMA system. Then, we formulate the minimization problem of the average weighted sum energy for view transmission and synthesis with respect to view selection and transmission power and subcarrier allocation. The problem is a challenging mixed discrete-continuous optimization problem with huge numbers of variables and constraints. Based on convex optimization and Difference of Convex (DC) programming, a low-complexity algorithm is proposed to obtain a suboptimal solution. Finally, advantages of the proposed suboptimal solution are numerically demonstrated.
\section{System Model}
\begin{figure}[!t]
	\centering
	\includegraphics[width=8.5cm,height=2.5cm]{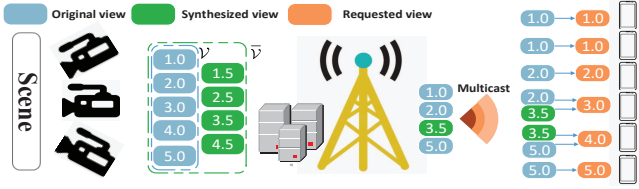}
	\caption{\small{Illustration of natural multicast and view synthesis-enabled multicast. Without view synthesis, the server can only exploit natural multicast opportunities, and has to transmit five views (i.e., 1.0, 2.0, 3.0, 4.0, 5.0). With view synthesis, the server can exploit both natural multicast opportunities and view synthesis-enabled multicast opportunities, and only needs to transmit four views (i.e., 1.0, 2.0, 3.5, 5.0) \cite{xu2019optimal}.}}
	
	%with $K=6$, $r_1=1$, $r_2=1$, $r_3=2$, $r_4=3$, $r_5=4$, $r_6=5$, $V=5$,  $\mathcal{V}=\{1,2,3,4,5\}$, $\overline{\mathcal{V}}=\{1,1.5,2,\cdots,5\}$, $\Delta_k=1$ for all $k\in\mathcal{K}$, $x_1=x_2=x_{3.5}=x_{5}=1$ and $y_{1,1}=y_{2,1}=y_{3,2}=y_{4,2}=y_{4,3.5}=y_{5,3.5}=y_{5,5}=y_{6,5}=1$.}}
	\label{system model}
	\vspace{-0.5cm}
\end{figure}

As shown in Fig.~\ref{system model}, a single-antenna server transmits an MVV to $K$ ($>$1) single-antenna users in an OFDMA system.\footnote{The MVV model and view selection model are the same as those in \cite{xu2019optimal}. We present the details here for completeness.} The set of user indices is denoted by $\mathcal{K}\triangleq\{1,\cdots,K\}$. There are $V$ original views, denoted by $\mathcal{V}\triangleq\{1,\cdots,V\}$. Between any two adjacent original views, there are $Q-1$ evenly spaced additional views, where $Q=2,3,\cdots$ is a system parameter. The additional views can be synthesized via DIBR. Let $\overline{\mathcal{V}}\triangleq\{1,1+1/Q,\cdots,V\}$ denote the set of indices for all views. Suppose that the source encoding rates of all views are the same, which is denoted by $R$ (in bits/s). 

The server stores all the original views in $\mathcal{V}$, and can synthesize any additional view $v$ in $\overline{\mathcal{V}} \setminus \mathcal{V}$ (if needed), based on its nearest left and right original views $\lfloor v \rfloor$ and $\lceil v \rceil$, where $\lfloor v \rfloor$ and $\lceil v \rceil$ denote the greatest integer less than or equal to $v$ and the least integer greater than or equal to $v$, respectively. Each user $k\in\mathcal{K}$ can synthesize any view $v\in \overline{\mathcal{V}} \setminus \{1,V\}$ based on two views in $\overline{\mathcal{V}}^-_{v,k} \triangleq \{x\in \overline{\mathcal{V}}: v-\Delta_k\leq x < v\}$ and $\overline{\mathcal{V}}^+_{v,k}\triangleq\{x\in\overline{\mathcal{V}}:v<x\leq v+\Delta_k\}$, respectively, which are successfully received by him. Here, $\Delta_k,k\in\mathcal{K}$ are system parameters, which reflect the qualities of synthesized views.

The view requested by user $k\in \mathcal{K}$, denoted by $r_k \in \overline{\mathcal{V}}$, is known at the server. When one view is requested by multiple users, natural multicast opportunities can be exploited to improve transmission efficiency. When different views are requested by multiple users, multicast opportunities may be created based on view synthesis, referred to as view synthesis-enabled multicast opportunities \cite{xu2019optimal}, to improve transmission efficiency. An illustration example can be found in Fig.~\ref{system model}.

We study a certain time duration, over which multiple groups of pictures (GOPs) are transmitted and each user's view angle does not change. The view transmission variable for view $v$ is denoted by:
\begin{equation}
x_v\in\{0,1\},\quad v\in \overline{\mathcal{V}}. \label{binary constraint x}
\end{equation}
Here, $x_v=1$ represents that view $v$ is transmitted by the server and $x_v=0$ otherwise. Denote $\mathbf{x} \triangleq (x_v)_{v \in \overline{\mathcal{V}}}$. The view utilization variable for view $v$ at user $k$ is denoted by:
\begin{equation}
y_{k,v}\in\{0,1\},\quad v\in \overline{\mathcal{V}},\ k \in \mathcal{K}. \label{binary constraint y}
\end{equation}
Here, $y_{k,v}=1$ represents that view $v$ is utilized by user $k$ and $y_{k,v}=0$ otherwise. Thus, to satisfy each user's view request, we require:
\begin{align}
&y_{k,r_k}+\sum\nolimits_{v\in \overline{\mathcal{V}}^-_{r_{k},k}} y_{k,v} = 1, \quad k\in \mathcal{K},  \label{Left constraint} \\
&y_{k,r_k}+\sum\nolimits_{v\in \overline{\mathcal{V}}^+_{r_{k},k}} y_{k,v} = 1, \quad k\in \mathcal{K}, \label{Right constraint} \\
&\sum\nolimits_{v\in \overline{\mathcal{V}} \setminus (\overline{\mathcal{V}}^-_{r_{k},k}\cup\overline{\mathcal{V}}^+_{r_{k},k})} y_{k,v}=0,\quad k\in\mathcal{K}. \label{rest zero contraint}
\end{align}
Each user can use only the transmitted views, i.e.,
\begin{equation}
	x_v \geq y_{k,v}, \quad k\in \mathcal{K},\ v\in \overline{\mathcal{V}}. \label{x>y}
\end{equation}
$(\mathbf{x},\mathbf{y})$ are referred to as view selection variables \cite{xu2019optimal}. Because of the coding structure of MVV, the values of $(\mathbf{x},\mathbf{y})$ cannot be changed during the considered time duration.

We consider an OFDMA system with $N$ subcarriers, denoted by $\mathcal{N}\triangleq\{1,\cdots,N\}$. The bandwidth of each subcarrier is $B$ (in Hz). Assume block fading and consider an arbitrary slot. Let $H_{n,k}\in \mathcal{H}$ denote the power of the channel for subcarrier $n$ and user $k$, where $\mathcal{H}$ denotes the finite channel state space. The system channel state is denoted by $\mathbf{H}\triangleq (H_{n,k})_{n\in\mathcal{N},k\in\mathcal{K}}\in  \boldsymbol{\mathcal{H}}\triangleq\mathcal{H}^{NK}$. The server is aware of the system channel state $\mathbf{H}$.

The subcarrier assignment indicator variable for subcarrier $n$ and view $v$ under the system channel state $\mathbf{h}\triangleq(h_{n,k})_{n\in\mathcal{N},k\in\mathcal{K}}\in \boldsymbol{\mathcal{H}}$ is denoted by:
\begin{equation}
\mu_{\mathbf{h},n,v} \in\{0,1\},\quad \mathbf{h}\in\boldsymbol{\mathcal{H}},~n\in\mathcal{N},~v\in\overline{\mathcal{V}}. \label{subcarrier indicator}
\end{equation}
Here, $\mu_{\mathbf{h},n,v}=1$ represents that the server uses subcarrier $n$ to transmit view $v$ under $\mathbf{h}$, and $\mu_{\mathbf{h},n,v}=0$ otherwise. Suppose that each subcarrier is allowed to transmit only one view. Thus, we have:
\begin{align}
&\sum_{v\in\overline{\mathcal{V}}} \mu_{\mathbf{h},n,v}=1, \quad \mathbf{h}\in\boldsymbol{\mathcal{H}},~n\in\mathcal{N}. \label{subcarrier constraints}
\end{align}

The transmission power and rate for view $v$ on subcarrier $n$ under $\mathbf{h}$ are denoted by $p_{\mathbf{h},n,v}$ and $c_{\mathbf{h},n,v}$, respectively, where
\begin{align}
&p_{\mathbf{h},n,v}\geq 0,\quad \mathbf{h}\in\boldsymbol{\mathcal{H}},~n\in\mathcal{N},~v\in\overline{\mathcal{V}}, \label{p>=0}\\
&c_{\mathbf{h},n,v}\geq 0,\quad \mathbf{h}\in\boldsymbol{\mathcal{H}},~n\in\mathcal{N},~v\in\overline{\mathcal{V}}. \label{c>=0}
\end{align}
Let $\mathbf{c}\triangleq (c_{\mathbf{h},n,v})_{\mathbf{h}\in\boldsymbol{\mathcal{H}},n\in\mathcal{N},v\in\overline{\mathcal{V}}}$. The total transmission energy consumption per time slot under $\mathbf{h}$ is given by:
\begin{equation*}
E(\boldsymbol{\mu}_\mathbf{h},\mathbf{p}_\mathbf{h})=T\sum_{n\in\mathcal{N}}\sum_{v\in\overline{\mathcal{V}}} \mu_{\mathbf{h},n,v}p_{\mathbf{h},n,v},
\end{equation*}
where $\boldsymbol{\mu}_\mathbf{h}\triangleq (\mu_{\mathbf{h},n,v})_{n\in\mathcal{N},v\in\overline{\mathcal{V}}}$ and $\mathbf{p}_{\mathbf{h}}\triangleq (p_{\mathbf{h},n,v})_{n\in\mathcal{N},v\in\overline{\mathcal{V}}}$. 
Capacity achieving code is considered to obtain design insights. To make sure that all users have no stall during the video playback, the following transmission rate constraints should be satisfied:
\begin{align}
&\sum_{n\in\mathcal{N}} c_{\mathbf{h},n,v}\geq x_v R, \quad  \mathbf{h}\in\boldsymbol{\mathcal{H}},v\in\overline{\mathcal{V}},\label{bandwidth constraints}\\
&B\mu_{\mathbf{h},n,v}\log_2\left(1+\frac{p_{\mathbf{h},n,v}h_{n,k}}{n_0}\right)\geq y_{k,v}c_{\mathbf{h},n,v},\nonumber\\ 
&\qquad\qquad\qquad \qquad\qquad\mathbf{h}\in\boldsymbol{\mathcal{H}},n\in\mathcal{N},k\in\mathcal{K},v\in\overline{\mathcal{V}},\label{each subcarrier rate}
\end{align}
where $n_0$ is the complex additive white Gaussian channel noise power for each subcarrier. Note that the communication model for transmission of an MVV to multiple users in an OFDMA system is more complicated than that in a TDMA system \cite{xu2019optimal}.

Let $E_b$ and $E_{\text{u},k}$ denote the energy consumption for synthesizing one view per time slot at the server and user $k$, respectively. Therefore, the weighted sum energy consumption per time slot under $\mathbf{h}$ is given by: $$E(\mathbf{x},\mathbf{y},\boldsymbol{\mu}_\mathbf{h},\mathbf{p}_\mathbf{h}) =E(\boldsymbol{\mu}_\mathbf{h},\mathbf{p}_\mathbf{h})  +E^{(b)}(\mathbf{x})+\beta E^{(u)}(\mathbf{y}).$$ 
Here, $\beta\geq 1$ is the corresponding weight factor for users, and $$E^{(b)}(\mathbf{x})\triangleq\sum_{v\in \overline{\mathcal{V}} \setminus \mathcal{V}}x_vE_b$$ and 
$$E^{(u)}(\mathbf{y})\triangleq\sum_{k\in \mathcal{K}} (1-y_{k,r_k})E_{\text{u},k}$$ denote the synthesis energy consumption per time slot at the server and all users, respectively. The average weighted sum energy consumption per time slot is given by:
$$\mathbb{E} [E(\mathbf{x},\mathbf{y},\boldsymbol{\mu}_\mathbf{H},\mathbf{p}_\mathbf{H})] =\mathbb{E}\left[E(\boldsymbol{\mu}_\mathbf{H},\mathbf{p}_\mathbf{H})\right]+E^{(b)}(\mathbf{x})+\beta E^{(u)}(\mathbf{y}),$$
where the expectation $\mathbb{E}$ is taken over $\mathbf{H}\in\boldsymbol{\mathcal{H}}$. 

\section{Problem Formulation and Structural Analysis}
In this section, the view selection, transmission power allocation and subcarrier allocation are optimized to minimize the average weighted sum energy consumption. In particular, consider the following optimization problem.
\begin{problem}[Energy Minimization]\label{View selection and resource allocation}
\begin{align}
	E^\star &\triangleq \min_{\mathbf{x},\mathbf{y},\boldsymbol{\mu},\mathbf{p},\mathbf{c}}\quad \mathbb{E} [E(\mathbf{x},\mathbf{y},\boldsymbol{\mu}_\mathbf{H},\mathbf{p}_\mathbf{H})]  \notag\\
	&\text{s.t.} ~~ (\ref{binary constraint x}),(\ref{binary constraint y}),(\ref{Left constraint}),(\ref{Right constraint}),(\ref{rest zero contraint}),(\ref{x>y}),(\ref{subcarrier indicator}),(\ref{subcarrier constraints}),(\ref{p>=0}),(\ref{c>=0}),(\ref{bandwidth constraints}),(\ref{each subcarrier rate}). \notag
\end{align}
\end{problem}
Let ($\mathbf{x}^\star,\mathbf{y}^\star,\boldsymbol{\mu}^\star,\mathbf{p}^\star,\mathbf{c}^\star$) denote an optimal solution of Problem~\ref{View selection and resource allocation}.
%where $\mathbf{x}^\star \triangleq (x_v^\star)_{v\in \overline{\mathcal{V}}}$, $\mathbf{y}^\star \triangleq (y^\star_{k,v})_{k\in \mathcal{K},v\in \overline{\mathcal{V}}}$, $\boldsymbol{\mu}^\star \triangleq (\mu_{\mathbf{h},n,v}^\star)_{\mathbf{h}\in\boldsymbol{\mathcal{H}},~n\in\mathcal{N},v\in\overline{\mathcal{V}}}$, $\mathbf{p}^\star \triangleq (p_{\mathbf{h},n,v}^\star)_{\mathbf{h}\in\boldsymbol{\mathcal{H}},~n\in\mathcal{N},v\in\overline{\mathcal{V}}}$ and $\mathbf{c}^\star\triangleq (c_{\mathbf{h},n,v}^\star)_{\mathbf{h}\in\boldsymbol{\mathcal{H}},n\in\mathcal{N},v\in\overline{\mathcal{V}}}$.

Problem~\ref{View selection and resource allocation} is a two-timescale optimization problem. Specifically, view selection is in a larger timescale and adapts to the channel distribution; power allocation and subcarrier allocation are in a shorter timescale and are adaptive to instantaneous channel powers. Problem~\ref{View selection and resource allocation} has $\mathcal{O}(N|\mathcal{H}|^{NK}+K)$ discrete variables, i.e., $(\mathbf{x},\mathbf{y},\boldsymbol{\mu})$ and $\mathcal{O}(N|\mathcal{H}|^{NK})$ continuous variables, i.e., $(\mathbf{p},\mathbf{c})$, as well as $\mathcal{O}(NK|\mathcal{H}|^{NK})$ constraints. There are two main challenges for solving Problem~\ref{View selection and resource allocation}: discrete variables are involved and the numbers of variables and constraints are huge. It is worth noting that due to the couping between view selection and resource allocation, traditional optimization techniques for subcarrier allocation cannot be directly applied to solve Problem~\ref{View selection and resource allocation}.

Define $\mathbf{Y}\triangleq\{\mathbf{y}:(\ref{binary constraint y}),(\ref{Left constraint}),(\ref{Right constraint}),(\ref{rest zero contraint})\}$. Introduce $P_{\mathbf{h},n,v}\triangleq p_{\mathbf{h},n,v}\mu_{\mathbf{h},n,v}$, eliminate $\mathbf{x}$ and $\mathbf{c}$, and separate $\mathbf{y}$ and $\boldsymbol{\mu},\mathbf{P}\triangleq(P_{\mathbf{h},n,v})_{\mathbf{h}\in\boldsymbol{\mathcal{H}},n\in\mathcal{N},v\in\overline{\mathcal{V}}}$. Then, we obtain a related problem, which will facilitate the optimization.
\begin{problem}[View Selection]\label{view selection}
	\begin{align}
	E^*\triangleq\min_{\mathbf{y}\in\mathbf{Y}}~ \mathbb{E} [E^*_{\mathbf{h}}(\mathbf{y})] + E_b\sum_{v\in \overline{\mathcal{V}} \setminus \mathcal{V}} \max_{k \in \mathcal{K}} y_{k,v}+\beta E^{(u)}(\mathbf{y}) \nonumber
	\end{align}
\end{problem}
where $E^*_{\mathbf{h}}(\mathbf{y})$ is given by the following subproblem.
\begin{problem}[Resource Allocation for $\mathbf{h}\in\boldsymbol{\mathcal{H}}$ and $\mathbf{y}\in\mathbf{Y}$]\label{subproblem}
	\begin{align}
	&E^*_{\mathbf{h}}(\mathbf{y}) \triangleq \min_{\boldsymbol{\mu}_\mathbf{h},\mathbf{P}_\mathbf{h}}\quad T\sum_{n\in\mathcal{N}} \sum_{v\in\overline{\mathcal{V}}} P_{\mathbf{h},n,v} \notag\\
	&\text{s.t.} \quad \mu_{\mathbf{h},n,v} \in\{0,1\},\quad n\in\mathcal{N},~v\in\overline{\mathcal{V}},\label{subproblem mu}\\
	&\qquad \sum_{v\in\overline{\mathcal{V}}} \mu_{\mathbf{h},n,v}=1, \quad n\in\mathcal{N},\label{subproblem sum mu}\\
	&\qquad P_{\mathbf{h},n,v}\geq 0,\quad n\in\mathcal{N},~v\in\overline{\mathcal{V}}, \label{subproblem P}\\
	&\qquad \sum_{n\in\mathcal{N}} B\mu_{\mathbf{h},n,v}\log_2\left(1+\frac{P_{\mathbf{h},n,v}H_{n,v}^{\min}(\mathbf{y}_v)}{\mu_{\mathbf{h},n,v}n_0}\right)\nonumber\\ 
	&\qquad\qquad\qquad\quad \geq R\max_{k \in \mathcal{K}} y_{k,v}, \quad n\in\mathcal{N},v\in\overline{\mathcal{V}}\label{subproblem each ch},
	\end{align}
\end{problem}
where $\mathbf{P}_\mathbf{h}\triangleq (P_{\mathbf{h},n,v})_{n\in\mathcal{N},v\in\overline{\mathcal{V}}}$, $\mathbf{y}_v\triangleq(y_{k,v})_{k\in\mathcal{K}}$ and $H_{n,v}^{\min}(\mathbf{y}_v)\triangleq\min\limits_{k\in\{k\in\mathcal{K}|y_{k,v}=1\}} H_{n,k}$.

Let $\mathbf{y}^*$ denote an optimal solution of Problem~\ref{view selection}. Let $(\boldsymbol{\mu}_\mathbf{h}^*(\mathbf{y}),\mathbf{P}_\mathbf{h}^*(\mathbf{y}))$ denote an optimal solution of Problem~\ref{subproblem}, where 
\begin{align*}
&\boldsymbol{\mu}_\mathbf{h}^*(\mathbf{y})\triangleq (\mu_{\mathbf{h},n,v}^*(\mathbf{y}))_{n\in\mathcal{N},v\in\overline{\mathcal{V}}},\\
&\mathbf{P}_\mathbf{h}^*(\mathbf{y})\triangleq (P_{\mathbf{h},n,v}^*(\mathbf{y}))_{n\in\mathcal{N},v\in\overline{\mathcal{V}}}.
\end{align*}
Denote 
\begin{align*}
&\mathbf{x}^*\triangleq (x_v^*)_{v\in\overline{\mathcal{V}}},\\
&\boldsymbol{\mu}^*(\mathbf{y}^*)\triangleq (\mu_{\mathbf{h},n,v}^*(\mathbf{y}^*))_{\mathbf{h}\in\boldsymbol{\mathcal{H}},n\in\mathcal{N},v\in\overline{\mathcal{V}}},\\
&\mathbf{p}^*(\mathbf{y}^*)\triangleq (p_{\mathbf{h},n,v}^*(\mathbf{y}^*))_{\mathbf{h}\in\boldsymbol{\mathcal{H}},n\in\mathcal{N},v\in\overline{\mathcal{V}}},\\
&\mathbf{c}^*(\mathbf{y}^*)\triangleq (c_{\mathbf{h},n,v}^*(\mathbf{y}^*))_{\mathbf{h}\in\boldsymbol{\mathcal{H}},n\in\mathcal{N},v\in\overline{\mathcal{V}}}.
\end{align*}
We have the following result.
\begin{lemma}\textit{(Relationship between Problems~\ref{view selection},\ref{subproblem} and Problem~\ref{View selection and resource allocation}):}\label{Optimal value property}
$E^\star=E^*$, $\mathbf{x}^\star=\mathbf{x}^*$, $\mathbf{y}^\star=\mathbf{y}^*$, $\boldsymbol{\mu}^\star=\boldsymbol{\mu}^*(\mathbf{y}^*)$, $\mathbf{p}^\star=\mathbf{p}^*(\mathbf{y}^*)$, and  $\mathbf{c}^\star=\mathbf{c}^*(\mathbf{y}^*)$,
where  
\begin{align*}
&x_v^*=\max_{k \in \mathcal{K}} y^*_{k,v},\quad  v\in\overline{\mathcal{V}},\\
&p_{\mathbf{h},n,v}^*(\mathbf{y}^*)=\frac{P_{\mathbf{h},n,v}^*(\mathbf{y}^*)}{\mu_{\mathbf{h},n,v}^*(\mathbf{y}^*)},\quad \mathbf{h}\in\boldsymbol{\mathcal{H}},~n\in\mathcal{N},~v\in\overline{\mathcal{V}},\\
&c_{\mathbf{h},n,v}^*(\mathbf{y}^*)=B\mu^*_{\mathbf{h},n,v}\log_2\left(1+\frac{P_{\mathbf{h},n,v}^*H_{n,v}^{\min}(\mathbf{y}^*_v)}{\mu^*_{\mathbf{h},n,v}n_0}\right),\\
&\qquad \qquad \qquad \qquad\qquad\qquad \mathbf{h}\in\boldsymbol{\mathcal{H}},~n\in\mathcal{N},~v\in\overline{\mathcal{V}}.
\end{align*}
\end{lemma}
\begin{IEEEproof}(sketch)
	First, by Lemma~2 in \cite{xu2019optimal}, we have $x_v^\star =\max\nolimits_{k\in\mathcal{K}} y_{k,v}^\star, v\in\overline{\mathcal{V}}$. Thus, we can eliminate $\mathbf{x}$ by replacing $x_v$ with $\max\nolimits_{k\in\mathcal{K}} y_{k,v}$, for all $ v\in\overline{\mathcal{V}}$. Then, following the proof for Lemma~1 in \cite{guo2018optimal}, we can eliminate $\mathbf{c}$ and simplify the constraints in (\ref{c>=0}), (\ref{bandwidth constraints}) and (\ref{each subcarrier rate}) to (\ref{subproblem each ch}) for all $\mathbf{h}\in\mathcal{H}$. Finally, using $P_{\mathbf{h},n,v}$ instead of $p_{\mathbf{h},n,v}$ and separating $\mathbf{y}$ and $\boldsymbol{\mu},\mathbf{P}$, we can obtain the equivalent formulation in Problem~\ref{view selection} and Problem~\ref{subproblem}.
\end{IEEEproof}

\section{Suboptimal Solution}
In this section, a low-complexity algorithm is proposed to obtain a suboptimal solution of Problem~\ref{View selection and resource allocation}, based on the equivalent formulation in Problem~\ref{view selection} and Problem~\ref{subproblem}.
\subsection{Optimal Solution of Problem~\ref{subproblem}} 
In this subsection, an optimal solution of Problem~\ref{subproblem} is obtained. By replacing the constraints in (\ref{subproblem mu}) with $\mu_{\mathbf{h},n,v}\geq 0, n\in\mathcal{N}, v\in\overline{\mathcal{V}}$, we can obtain the relaxed version of Problem~\ref{subproblem}, which is convex and can be easily solved. By Lemma~1 in \cite{guo2018optimal}, we know that under a mild condition,\footnote{The condition can be easily satisfied when the channel conditions of users vary from each other \cite{guo2018optimal}.} the optimal solution of the relaxed convex problem of Problem~\ref{subproblem} is also an optimal solution of Problem~\ref{subproblem}. Note that in previous works on MVV transmission in OFDMA systems, optimal power and subcarrier allocation has not been obtained.

\begin{comment}
\begin{lemma}[Optimal Solution of Problem~\ref{subproblem}]
	Suppose that for all $n\in\mathcal{N}$, there exists a unique $v_n$ such that $W_{n,v_n}(\mathbf{y}_{v_n},\lambda_{v_n}) = \max_{v\in\overline{\mathcal{V}}} W_{n,v}(\mathbf{y}_{v},\lambda_{v})$. Then, an optimal solution of Problem~\ref{subproblem} is $(\boldsymbol{\mu}_\mathbf{h}^*(\mathbf{y}),\mathbf{p}_\mathbf{h}^*(\mathbf{y}),\mathbf{c}_\mathbf{h}^*(\mathbf{y}))$ , where for all $n\in\mathcal{N}$ and $v\in\overline{\mathcal{V}}$,
	\begin{align*}
	&\mu_{\mathbf{h},n,v}^*(\mathbf{y}_{v}) = 
	\begin{cases}
	1, & v = \argmax_{v\in\overline{\mathcal{V}}}W_{n,v}(\mathbf{y}_{v},\lambda_{v}^*(\mathbf{y}_v)),\\
	0, & \text{otherwise},
	\end{cases}\\
	&p_{\mathbf{h},n,v}^*(\mathbf{y}_{v})=\mu_{\mathbf{h},n,v}^*(\mathbf{y}_{v})f_{n,v}(\mathbf{y}_v,\lambda_v^*(\mathbf{y}_{v}))\\
	&c_{\mathbf{h},n,v}^*(\mathbf{y}_{v})=B\log_2\left(1+\frac{p_{\mathbf{h},n,v}^*(\mathbf{y}_{v})H_{n,v}^{\min}(\mathbf{y}_v)}{n_0}\right)
	\end{align*} 
	Here, $\lambda_{v}^*(\mathbf{y}_v)$ satisfies
	\begin{align*}
	&\sum\nolimits_{n\in\mathcal{N}} \mu_{\mathbf{h},n,v}^*(\mathbf{y}_{v}) B\log_2\left(1+\frac{p_{\mathbf{h},n,v}^*(\mathbf{y}_{v})H_{n,v}^{\min}(\mathbf{y}_v)}{n_0}\right)\\
	&=R\max\nolimits_{k\in\mathcal{K}} y_{k,v}
	\end{align*}
\end{lemma}

$\lambda_{v}^*(\mathbf{y}_v),v\in\overline{\mathcal{V}}$ can be obtained using the subgradient method.
\end{comment}
\subsection{Approximate Solution of Problem~\ref{view selection}}
Obtaining an optimal solution of Problem~\ref{view selection} using exhaustive search is not acceptable, when $K$ and $N$ are large. In this subsection, a low-complexity approximate solution of Problem~\ref{view selection} is proposed. Similarly, we use $P_{\mathbf{h},n,v}$ instead of $p_{\mathbf{h},n,v}$ for all $\mathbf{h}\in\boldsymbol{\mathcal{H}},n\in\mathcal{N},v\in\overline{\mathcal{V}}$ and eliminate $\mathbf{x}$. In addition, we replace (\ref{bandwidth constraints}) and (\ref{each subcarrier rate}) with
\begin{align}
&\sum_{n\in\mathcal{N}} \bar{c}_{n,v}\geq R\max_{k \in \mathcal{K}} y_{k,v} , \quad v\in\overline{\mathcal{V}} \label{new bandwidth constraints}\\
&B \bar{\mu}_{n,v}\log_2\left(1+\frac{\bar{P}_{n,v}\bar{H}_k}{\bar{\mu}_{n,v}n_0}\right)\geq y_{k,v}\bar{c}_{n,v},\nonumber\\
&\qquad \qquad \quad \qquad \quad \quad\qquad\qquad n\in\mathcal{N},k\in\mathcal{K},v\in\overline{\mathcal{V}}. \label{new each subcarrier rate}
\end{align}
Here, $\bar{P}_{n,v}$, $\bar{\mu}_{n,v}$ and $\bar{c}_{n,v}$ approximately characterize $\mathbb{E}[P_{\mathbf{H},n,v}]$, $\mathbb{E}[\mu_{\mathbf{H},n,v}]$ and $\mathbb{E}[c_{\mathbf{H},n,v}]$, and $\bar{H}_k\triangleq\mathbb{E}[H_{n,k}]$. The goal is to reduce the numbers of variables and constraints by imposing approximate constraints on the average resource allocation.  Let $\bar{\mathbf{P}}\triangleq (\bar{P}_{n,v})_{n\in\mathcal{N},v\in\overline{\mathcal{V}}}$,  $\bar{\boldsymbol{\mu}}\triangleq (\bar{\mu}_{n,v})_{n\in\mathcal{N},v\in\overline{\mathcal{V}}}$ and $\bar{\mathbf{c}}\triangleq (c_{n,v})_{n\in\mathcal{N},v\in\overline{\mathcal{V}}}$. Therefore, we can obtain the following approximate problem.
\begin{problem}[Approximation of Problem~\ref{View selection and resource allocation}]\label{approximate problem}
	\begin{align}
	&\min_{\mathbf{y},\bar{\mathbf{P}},\bar{\boldsymbol{\mu}},\bar{\mathbf{c}}}\quad T\sum_{n\in\mathcal{N}} \sum_{v\in \overline{\mathcal{V}}} \bar{P}_{n,v} + E_b\sum_{v\in \overline{\mathcal{V}} \setminus \mathcal{V}} \max_{k \in \mathcal{K}} y_{k,v} + \beta E^{(u)}(\mathbf{y})  \notag\\
	&\text{s.t.} \quad (\ref{binary constraint y}),(\ref{Left constraint}),(\ref{Right constraint}),(\ref{rest zero contraint}),(\ref{new bandwidth constraints}),(\ref{new each subcarrier rate}), \nonumber\\
	&\qquad  \bar{\mu}_{n,v} \geq 0,\quad n\in\mathcal{N}, v\in\overline{\mathcal{V}}, \label{U}\\
	&\qquad \sum_{n\in\mathcal{N}} \bar{\mu}_{n,v}=1,\quad  v\in\overline{\mathcal{V}},\label{U constraints}\\
	&\qquad  \bar{c}_{n,v} \geq 0 ,\quad n\in\mathcal{N}, v\in\overline{\mathcal{V}}. \label{C}
	\end{align}
\end{problem}
Let $(\bar{\mathbf{y}}^\star,\bar{\mathbf{P}}^\star,\bar{\boldsymbol{\mu}}^\star,\bar{\mathbf{c}}^\star)$ denote an optimal solution of Problem~\ref{approximate problem}.

To further reduce computational complexity, we can obtain an equivalent problem of Problem~\ref{approximate problem} with a smaller number of variables by carefully exploring structural properties of Problem~\ref{approximate problem}.
\begin{problem}[Equivalent Problem of Problem~\ref{approximate problem}]\label{equivalent approximate problem}
	\begin{align}
	&\min_{\mathbf{y},\mathbf{L}}\quad n_0T\sum_{v\in \overline{\mathcal{V}}}L_v \max_{k \in \mathcal{K}} \left\{\frac{ 1}{\bar{H}_k}\left(2^{\frac{y_{k,v}R}{L_v B}}-1\right) \right\} \nonumber \\
	&\qquad \qquad  + E_b\sum_{v\in \overline{\mathcal{V}} \setminus \mathcal{V}} \max_{k \in \mathcal{K}} y_{k,v} + \beta E^{(u)}(\mathbf{y})+ \rho P(\mathbf{y})  \notag\\
	&\text{s.t.} \quad (\ref{Left constraint}),(\ref{Right constraint}),(\ref{rest zero contraint}), \nonumber\\
	&\qquad  L_v \geq 0,\quad v\in\overline{\mathcal{V}},\label{subcarrier number}\\
	&\qquad  \sum_{v\in\overline{\mathcal{V}}} L_{v}\leq N,\label{subcarrier number constraints}\\
	&\qquad y_{k,v}\in [0,1], \quad k\in\mathcal{K},~v\in \overline{\mathcal{V}}, \label{relaxed y}
	\end{align}
\end{problem}
where $\mathbf{L}\triangleq(L_v)_{v\in\overline{\mathcal{V}}}$, $L_v$ represents the number of subcarriers allocated for the transmission of view $v$, $\rho>0$ is the penalty parameter, the penalty function $P(\mathbf{y})$ is given by: $$P(\mathbf{y})=\sum_{k\in \mathcal{K}}\sum_{v\in \overline{\mathcal{V}}} y_{k,v}(1-y_{k,v}).$$ The optimal solution of Problem~\ref{equivalent approximate problem} is denoted by $(\bar{\mathbf{y}}^*,\bar{\mathbf{L}}^*)$.
\begin{lemma}[Equivalence Between Problems~\ref{approximate problem} and \ref{equivalent approximate problem}]\label{DC lemma}
There exists $\rho_0>0$ such that for all $\rho>\rho_0$, $\bar{\mathbf{y}}^\star=\bar{\mathbf{y}}^*$.  
\end{lemma}
\begin{IEEEproof}(sketch)
	First, we can eliminate $\bar{\mathbf{c}}$ and simplify the constraints in (\ref{new bandwidth constraints}), (\ref{new each subcarrier rate}) and (\ref{C}) to $$\sum_{n\in\mathcal{N}} B \bar{\mu}_{n,v}\log_2\left(1+\frac{\bar{P}_{n,v}\bar{H}_k}{\bar{\mu}_{n,v}n_0}\right) \geq y_{k,v}R,~k\in\mathcal{K},v\in\overline{\mathcal{V}}.$$ Then, following the proof of Lemma~3 in \cite{guo2018optimal}, we have 
	\begin{equation*}
	\bar{p}_{n,v}^\star=\begin{cases}
	\max\limits_{k \in \mathcal{K}}\left\{\frac{1}{\bar{H}_k}\left(2^{\frac{y_{k,v}^\star R}{L_v^\star B}}-1\right)\right\},&\mu_{n,v}^\star=1,\\
	0,&\mu_{n,v}^\star=0.
	\end{cases}
	\end{equation*}
	Thus, we can replace $T\sum_{n\in\mathcal{N}} \sum_{v\in \overline{\mathcal{V}}} \bar{P}_{n,v}$ in the objective function of Problem~\ref{approximate problem} with $$n_0T\sum_{v\in \overline{\mathcal{V}}}L_v \max_{k \in \mathcal{K}} \left\{ \frac{ 1}{\bar{H}_k}(2^{\frac{y_{k,v}R}{L_v B}}-1)\right\}.$$ Finally, by Theorem 8 of \cite{le2012exact}, we complete the proof.
\end{IEEEproof}

Based on Lemma~\ref{DC lemma}, we can solve Problem~\ref{equivalent approximate problem} (for a sufficiently large $\rho>0$) instead. As $P(\mathbf{y})$ is concave and (\ref{Left constraint})-(\ref{rest zero contraint}), (\ref{subcarrier number})-(\ref{relaxed y}) are convex, Problem~\ref{equivalent approximate problem} is a penalized DC problem. Thus, a stationary point of Problem~\ref{equivalent approximate problem} can be obtained by using the DC algorithm. Specifically, a sequence of convex approximations of Problem~\ref{equivalent approximate problem}, each of which is obtained by linearizing the penalty function $P(\mathbf{y})$, are solved iteratively~\cite{le2012exact}. Note that the convex approximation of Problem~\ref{equivalent approximate problem} has $\mathcal{O}(K)$ variables and $\mathcal{O}(K)$ constraints, which are much smaller than those of Problem~\ref{View selection and resource allocation}. The DC algorithm can be conducted multiple times, with random initial feasible points of Problem~\ref{equivalent approximate problem}, and the stationary point of Problem~\ref{equivalent approximate problem} that achieves the minimum average weighted sum energy consumption and zero penalty, denoted by $\bar{\mathbf{y}}^\dagger$, will be selected.

\subsection{Low-Complexity Algorithm for Problem~\ref{View selection and resource allocation}}
In this subsection, a low-complexity algorithm is proposed to obtain a suboptimal solution of Problem~\ref{View selection and resource allocation}. First, we obtain $\bar{\mathbf{y}}^\dagger$ using the DC algorithm for Problem~\ref{equivalent approximate problem}. Then, given $\bar{\mathbf{y}}^\dagger$, we solve Problem~\ref{subproblem} for all $\mathbf{h}\in\boldsymbol{\mathcal{H}}$ in parallel. Algorithm~\ref{DC Algorithm} shows the details. $(\bar{\mathbf{x}}^\dagger(\bar{\mathbf{y}}^\dagger),\bar{\mathbf{y}}^\dagger,\boldsymbol{\mu}^*(\bar{\mathbf{y}}^\dagger),\mathbf{p}^*(\bar{\mathbf{y}}^\dagger),\mathbf{c}^*(\bar{\mathbf{y}}^\dagger))$ can be regarded as a suboptimal solution of Problem~\ref{View selection and resource allocation}.
\begin{algorithm}[t]
	\small
	\caption{Obtaining A Suboptimal Solution of Problem~\ref{View selection and resource allocation}}
	\mbox{\textbf{Output} $(\bar{\mathbf{x}}^\dagger(\bar{\mathbf{y}}^\dagger),\bar{\mathbf{y}}^\dagger,\boldsymbol{\mu}^*(\bar{\mathbf{y}}^\dagger),\mathbf{p}^*(\bar{\mathbf{y}}^\dagger),\mathbf{c}^*(\bar{\mathbf{y}}^\dagger))$}
	\vspace{-0.2cm}
	\begin{algorithmic}[1] \label{DC Algorithm}
		\STATE Obtain $\bar{\mathbf{y}}^\dagger$ by solving Problem~\ref{equivalent approximate problem} using the DC algorithm.
		\STATE Set $\bar{\mathbf{x}}^\dagger(\bar{\mathbf{y}}^\dagger)\triangleq(x_v^\dagger(\bar{\mathbf{y}}^\dagger))_{v\in\overline{\mathcal{V}}}$, where $x_v^\dagger(\bar{\mathbf{y}}^\dagger) = \max\limits_{k \in \mathcal{K}} y^\dagger_{k,v}$, $v \in \overline{\mathcal{V}}$.
		\FOR{$\mathbf{h} \in \boldsymbol{\mathcal{H}}$}
		\STATE Obtain $(\boldsymbol{\mu}^*_\mathbf{h}(\bar{\mathbf{y}}^\dagger),\mathbf{p}^*_\mathbf{h}(\bar{\mathbf{y}}^\dagger),\mathbf{c}^*_\mathbf{h}(\bar{\mathbf{y}}^\dagger))$ by solving the relaxed convex problem of Problem~\ref{subproblem} with standard convex optimization techniques.
		\ENDFOR
	\end{algorithmic}
\end{algorithm}
% We use MVV sequence \emph{Kendo} as the video source and use HEVC in FFmpeg to encode the video with quantization parameter 15, frame rate 30 frame/s and resolution 1024$\times$768
\section{Numerical Results}
In this section, we numerically compare the proposed solution with two baseline schemes. In Baseline~1, view synthesis is adopted at the server but is not adopted at each user\cite{zhao2015qos}. In Baseline~2, view synthesis is adopted at each user but is not adopted at the server\cite{zhang2018packetization}. Note that natural multicast opportunities can be utilized by both baseline schemes; Baseline~1 cannot create multicast opportunities based on view synthesis, but can ensure that no more than $K$ views are transmitted; Baseline~2 can create multicast opportunities based on view synthesis, but may transmit more than $K$ views. Both baseline schemes adopt optimal transmission power and subcarrier allocation using the method in \cite{guo2018optimal}.  We adopt FFmpeg as the MVV encoder and MVV sequence \emph{Kendo} as the video source. For all $n\in\mathcal{N},k\in\mathcal{K}$, assume $H_{n,k}\sim \text{Exp}(10^{6})$. Suppose that views are randomly requested by the $K$ users in an i.i.d. manner, as in \cite{xu2019optimal}.\footnote{We omit the detailed view request model due to page limitation. Please refer to \cite{xu2019optimal} for details.} 100 realizations of $r_k, k\in\mathcal{K}$ and 500 realizations of $\mathbf{h}$ are randomly generated. The average performance is evaluated.
\begin{figure}[!t]
	\vspace{-0.5cm}
	\begin{center}
	\subfigure[Weighted sum energy versus $K$ with $\gamma=0.5$.]{ %第一张子图
		\begin{minipage}{0.61\linewidth}
			\centering
			\includegraphics[width=0.9\linewidth]{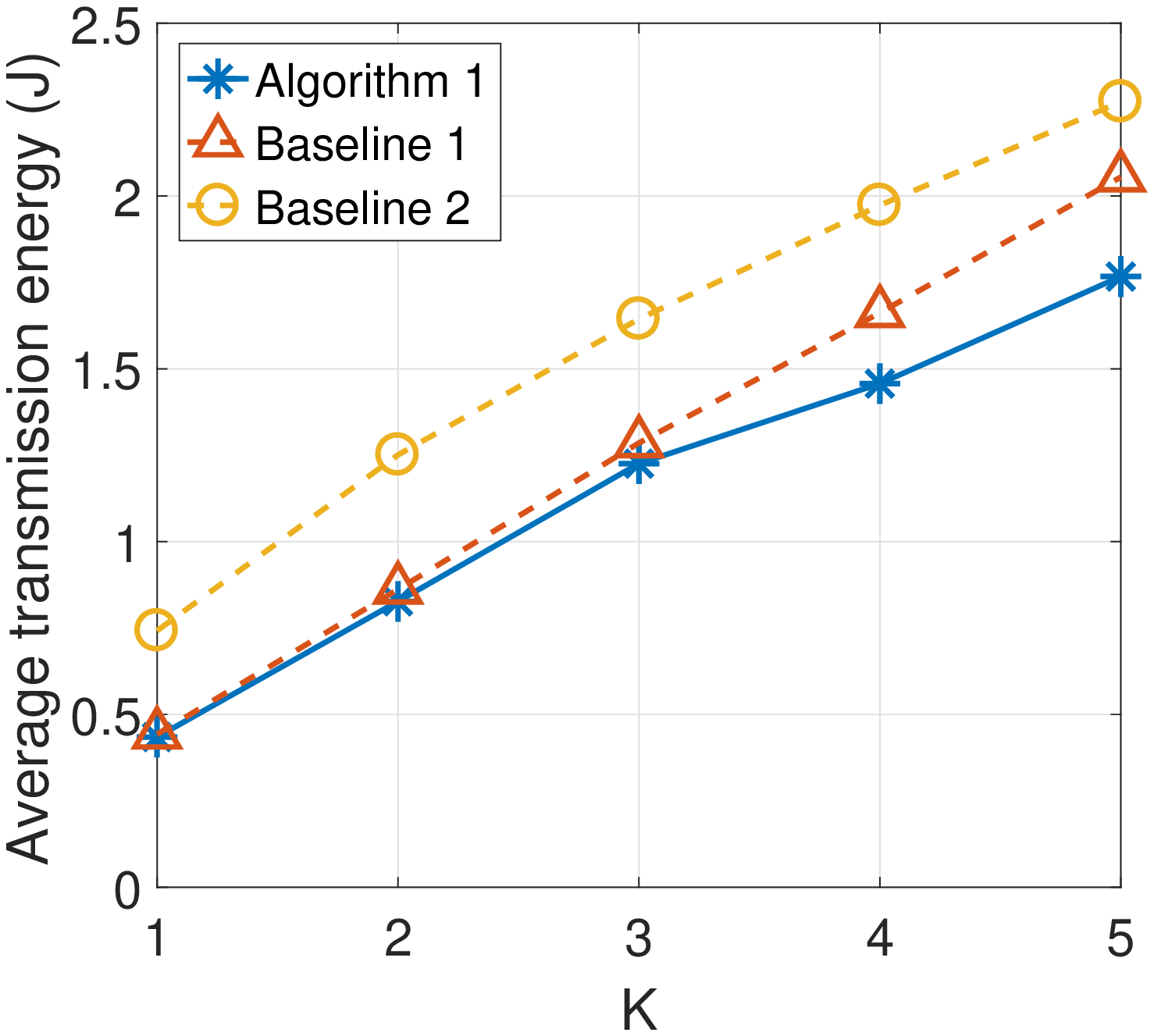}
		\end{minipage}
	}
	\subfigure[Weighted sum energy versus $\gamma$ with $K=5$.]{
		\begin{minipage}{0.61\linewidth}
			\centering
			\includegraphics[width=0.9\linewidth]{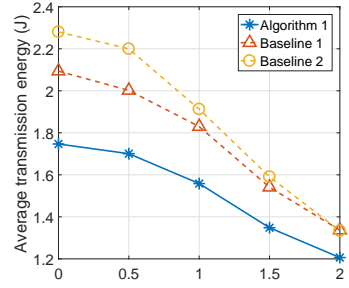}
		\end{minipage}
	}
	\caption{\small{Performance comparison. $R=18.59$~Mbit/s, $\beta=2$,  $E_b=10^{-3}$~Joule, $E_{\text{u},k}=10^{-3}$ Joule, $k \in \mathcal{K}$, $B=312.5$~KHz, $\mathcal{V}=\{1,2,3,4,5\}$,  $N=64$, $T=100$~ms, $Q=5$ and $n_0=10^{-9}$W.}}
	\end{center}
\vspace{-0.5cm}
\end{figure}

Fig.~2 illustrates the weighted sum energy consumption versus the number of users $K$ and the Zipf exponent $\gamma$. Fig.~2~(a) shows that as $K$ increases, the weighted sum energy consumption of each scheme increases, due to the load increase. Fig.~2~(b) shows that as $\gamma$ increases (i.e., view requests from the users are more concentrated), the weighted sum energy consumption of each scheme decreases, due to the increment of natural multicast opportunities. From Fig.~2, we can see that Baseline~1 outperforms Baseline~2, which reveals that naive creation of multicast opportunities usually causes extra transmission and yields a higher energy consumption. In addition, we can see that Algorithm~1 outperforms both baseline schemes, indicating the importance of the optimization of view synthesis-enabled multicast opportunities. The gains of Algorithm~1 over the two baseline schemes are significant at large $K$ or small $\gamma$, as more view synthesis-enabled multicast opportunities can be created.
\section{Conclusion and Future Works}
In this letter, we studied the transmission of an MVV to multiple users in an OFDMA system. We exploited both natural multicast opportunities and view synthesis-enabled multicast opportunities to improve transmission efficiency. First, we established a communication model for transmission of an MVV to multiple users in an OFDMA system. Then, we optimized view selection, transmission power and subcarrier allocation to minimize the average weighted sum energy consumption. A low-complexity algorithm was proposed to obtain a suboptimal solution of the challenging problem. The proposed optimization method can be readily extended to tackle two-timescale optimal resource allocation problems in OFDMA systems. This paper opens up several directions for future research. For instance, the proposed multicast mechanism and optimization framework can be extended to design optimal multi-quality MVV transmission in OFDMA systems. In addition, a possible direction for future research is to design optimal MVV transmission in different wireless systems.

%\begin{thebibliography}{1}

%\bibitem{}
%Thomas Lipp1?Stephen Boyd, "Variations and extension of the convex-concave procedure"

%\end{thebibliography}

%\bibliography{draft9} 

\end{document}